\documentclass[12pt,preprint]{elsarticle}
\def\lesssim{\mathrel{\hbox{\rlap{\hbox{\lower4pt\hbox{$\sim$}}}\hbox{$<$}}}}
\def\gtrsim{\mathrel{\hbox{\rlap{\hbox{\lower4pt\hbox{$\sim$}}}\hbox{$>$}}}}

\def\sm{M_\odot}

\def\B{\begin{equation}}
\def\E{\end{equation}}

\def\B{\begin{equation}}
\def\E{\end{equation}}

\usepackage{graphicx}

\usepackage{glossaries}

\newglossaryentry{key}
	{
		name={},
		description={},
		text={},
		first={}
	}
		
\newglossaryentry{GRB}
	{
		name={Gamma Ray Bursts},
		description={none},
		text={GRB},
		first={gamma ray bursts (GRB)}
	}

\newglossaryentry{LKP}
	{
		name={Lightest Kaluza-Klein Particle},
		description={},
		text={LKP},
		first={lightest first level KK particle (LKP)}
	}

\newglossaryentry{KK}
	{
		name={Kaluza-Klein},
		description={},
		text={KK},
		first={Kaluza-Klein (KK)}
	}
		
\newglossaryentry{SM}
	{
		name={standard model},
		description={},
		text={SM},
		first={standard model (SM)}
	}
		
\newglossaryentry{LUX}
	{
		name={LUX},
		description={},
		text={LUX},
		first={Large Underground Xenon (LUX)},
		long={Large Underground Xenon}
	}
		
\newglossaryentry{DM}
	{
		name={dark matter},
		description={},
		text={DM},
		first={dark matter (DM)}
	}
		
\newglossaryentry{AGN}
	{
		name={active galactic nuclei},
		description={},
		text={AGN},
		first={active galactic nuclei (AGN)}
	}
		
\newglossaryentry{UHECR}
	{
		name={ultra-high energy cosmic rays},
		description={},
		text={UHECR},
		first={ultra-high energy cosmic rays}
	}
		
\newglossaryentry{LHC}
	{
		name={Large Hadron Collider},
		description={},
		text={LHC},
		first={Large Hadron Collider (LHC)}
	}
		
\newglossaryentry{ISM}
	{
		name={interstellar medium},
		description={},
		text={ISM},
		first={interstellar medium (ISM)}
	}
		
\newglossaryentry{HAWC}
	{
		name={High-Altitude Water Cherenkov Observatory},
		description={},
		text={HAWC},
		first={High-Altitude Water Cherenkov Observatory (HAWC)}
	}
		
\newglossaryentry{CTA}
	{
		name={Cherenkov Telescope Array},
		description={},
		text={CTA},
		first={Cherenkov Telescope Array (CTA)}
	}
		
\newglossaryentry{SNR}
	{
		name={supernova remnants},
		description={},
		text={SNR},
		first={supernova remnants (SNR)}
	}

\makeglossaries

\begin{document}

\begin{frontmatter}
	\title{Gamma Ray Bursts as Sources of Exotic Particles}


	\author[smcm]{Ian Morgan}
		\cortext[corresponding]{Corresponding author}
		\ead{ilmorgan@smcm.edu}
	\author[smcm]{Ted Tao\corref{corresponding}}
		\ead{ttao@smcm.edu}
	\author[smcm]{Erin De Pree}
		\ead{ekdepree@smcm.edu} 
	\author[smcm,osu]{Kevin Tennyson}
	
	\address[smcm]{Department of Physics, St. Mary's College of Maryland, St. Mary's City MD 20686}
	\address[osu]{College of Earth, Ocean, and Atmospheric Sciences, Oregon State University, Corvallis, OR  97331}

	\begin{abstract}
	
		We consider the possible production of stable \gls{LKP} in baryonic \gls{GRB} out flows. We numerically computed the energy-dependent cross-sections 
		of \gls{KK} excitations for the Standard Model gauge bosons, $\gamma$ and $Z^{0}$. Next, we determined the feasibility of producing these \gls{KK} 
		excitations in gamma-ray emitting regions of \gls{GRB}s. We found that a \gls{GRB} fireball that accelerates baryons to energies greater than $10^{14}\rm$ eV 
		could produce \gls{KK} excitations out to approximately $10^{12}\, \rm cm$ from the central engine, indicating that \gls{GRB}s may be a significant source of 
		the \gls{LKP}. Finally, we explore the potential observational consequences of our results.
		
	\end{abstract}
	
	\begin{keyword}
		gamma-ray burst \sep kaluza-klein \sep dark matter candidate
	\end{keyword}

\end{frontmatter}

\glsresetall

\section{Introduction}
	
	\gls{KK} theory allows all \gls{SM} fields to propagate into compact extra dimensions \citep{ss82}, which is a simple extra-dimensional 
	extension of the \gls{SM} with a dark matter candidate. All fields are assumed to propagate in a flat space-time $\mathcal{M}_4 \times \mathcal{S}_1$  
	where $\mathcal{S}_1$ is the extra dimension compactified on a circle.  Since the fields are confined and must be continuous, a discrete series of particles arises.  
	The momentum of the particles in the extra dimension appears as part of the apparent 4D mass.  This creates a tower of \gls{KK} resonances (or excitations) for each 
	particle.  Each excitation is heavier than the last and has the same quantum numbers and spin as the 4D particle.
	
	The mass of the $n$th mode in the \gls{KK} tower, $m^n$, is given by,
	\B
		(m^n)^2 = \left( \frac{n}{R} \right)^2 + (m^0)^2.
	\E
	where $m^0$ is the mass of the 4D particle (also called the zeroth order \gls{KK} mode) and $R$ is the radius of the compactified extra dimension.
	
	The momentum conservation in the extra dimensions leads to the conservation of \gls{KK} number at each vertex. Conservation of \gls{KK} parity requires that 
	interactions only occur between even-number or odd-number \gls{KK} modes. Thus, there exists a \gls{LKP}, usually a linear combination the \gls{KK} photon and 
	the \gls{KK} with suppressed decay cross section to the \gls{SM} zero modes \citep{2010JHEP...09..025B}. The unstable \gls{KK} modes will quickly decay into 
	the \gls{SM} zero mode or the stable \gls{LKP}. The identity of the \gls{LKP} crucially depends on the mass spectrum of the first \gls{KK} mode. Moreover, the 
	\gls{LKP} is considered a promising dark matter candidate because it is long-lived and has low annihilation cross sections \citep{2003NuPhB.650..391S}. The 
	current \gls{LHC} limit on the compactification scale is $R^{-1} \geq 715$ GeV and the mass of the \gls{LKP} $M_{KK}\geq 1.4 $ TeV \citep{2013JHEP...08..091E}.  
	
	Experimentally, direct detection searches are attempt to isolate rare interactions between \gls{DM} and ordinary matter. For example, the \gls{LUX} 
	\cite{2012NIMPA.668....1A} and XENON-100 \citep{2012APh....35..573X} dark matter experiments searched for events in liquid xenon target by recreating the light 
	received by photomultipliers. These efforts are limited by a \gls{DM} candidate's mass and its scattering cross-section, and their maximum possible sensitivity depends 
	on background signal from coherent neutrino scattering.
	
	There are also indirect methods that search for evidence of rare annihilation or scattering events elsewhere in the universe. In fact, the experimental results from 
	numerous instruments, e.g.~Fermi and PAMELA, have noted an excess in the cosmic electron-positron spectrum, which may be consistent with the annihilation 
	events of 3 TeV \gls{DM} \citep{2009APh....32..140G, 2009NJPh...11j5023B}. However, these findings are open to interpretation, especially since the signals cannot 
	be localized to known \gls{DM} sources \citep{2009PhRvL.103e1104B}. 
	
	On the theoretical front, recent work on indirect astrophysical detection of exotic dark matter candidates \citep{bw98} explored radiation signatures from 
	electron-dark matter scattering near active galactic nuclei, reaching the rather pessimistic conclusion that such signals are probably undetectable with existing and 
	near future instruments. In the more specific context of \gls{AGN} jets, \cite{gpu10} performed a more detailed calculation that included specific candidate particles 
	including the \gls{KK}. They found that resonances dictated by exotic particle mass spectrum may dominate scattering between proton (or electron) and dark matter 
	candidates, resulting in a potentially detectable break in the gamma ray spectra from \gls{AGN}s. However, these studies assumed that the exotic dark matter already 
	exist in sufficient quantities, and did not consider how such particles may have been produced.
	
	In this work, we propose a potential astrophysical scenario for the production of the \gls{LKP} as \gls{DM}, and discuss the possible observational implication. 
	\gls{DM} searches that rely on rare interactions, restricts the number of events that we can observe. Particle accelerators overcome this by creating events, yet 
	are limited by the energies that are required to create these events. However, some cosmic particle accelerators in our universe that do not have the same energy 
	constraints, including \glsname{AGN}, \gls{SNR}, and \gls{GRB}. The improving sensitivity of gamma-ray instruments may enable us to search for potential \gls{DM} 
	matter signatures in the gamma ray regime from such cosmic accelerators.
	
	We show that a combination of the current theories for \gls{KK} \gls{DM} and \gls{GRB} dynamics means that \gls{GRB}s could produce \gls{KK} \gls{DM} in 
	appreciable quantities. \gls{GRB}s are attractive possibilities because they are, at least in principle, plausible sources for \gls{UHECR} at energies between 
	$10^{18.5}$ to $10^{20.5}\rm\, eV$, especially if the lower energy cosmic rays can come from other systems such as supernova remnants \citep{wax95, vietri95}. 
	In this context, baryon-loaded \gls{GRB} fireballs may accelerate protons and neutrons to above $10^{20} \rm\, eV$ \citep{wax06}, possibly via a combination of 
	first and second order Fermi acceleration \citep{mur12} at shocks. Furthermore, \gls{GRB}s create mildly relativistic shocks with the \gls{ISM} 
	\citep{1949PhRv...75.1169F}. They may also create strongly shocked regions, when fireball outflows, with randomly distributed energies, collide. Moreover, the 
	millisecond variability of \gls{GRB} light curves may rise from multiple shells ejected by the central engine, so it is possible for there to be several internal collisions. 
	Particle encounters about these shocks generate particle showers which are translated into the gamma-rays. These particle showers may be frequent and energetic 
	enough to produce \gls{KK} excitations. Beyond high energy cosmic rays physics, these hadronic models also have garnered significant attention due to their potential 
	to explain a range of observations \citep{aim09, agm09, ra08}.  We refer interested readers to review articles such as \cite{grf09, gr13, mes13} and \citep{ber14} as 
	well as further references for more details regarding the rich and fascinating phenomenology of \gls{GRB}s.

	We use the fireball model from \citep{zhang02} to estimate physical properties of the fireball. We assume a baryon-dominated fireball as such outflows are more 
	likely to generate significant quantities of high-energy particle encounters compared to Poynting dominated scenarios. Next, we numerically compute the 
	energy-averaged cross-section of processes involving \gls{KK} particles, and combine these results with the collision region parameters to obtain an excitation rate 
	and optical depth. These results then enable us to identify, at a qualitative level, potential signatures of the \gls{KK} theory in \gls{GRB}s.
	
	Our paper is organized as follows. In Section \ref{sec:xsection} we outline our numerical cross-section computations and some necessary aspects of \gls{KK} particle 
	physics. Section \ref{sec:dynamics} then explains the \gls{GRB} dynamics relevant to our calculation while summarizing relevant results from \cite{zhang02}. Section 
	\ref{sec:collisions} the explains the reaction rate calculations. In Section \ref{sec:results} we present our main findings. In Section \ref{sec:conclusion}, we discuss 
	our results and the potential observational implications.

\section{Methods}
	\subsection{KK excitation at high energies}
	\label{sec:xsection}
		We used the computer program Pythia 8 \citep{2008CoPhC.178..852S} to simulate proton-proton collisions with and without a TeV$^{-1}$ sized extra 
		dimension \citep{2010JHEP...09..025B}. We assume that any resulting \gls{KK} tower quickly decays into either the \gls{LKP} or \gls{SM} particles. The 
		\gls{LKP} was given a set mass of 1.5 TeV, slightly above the current limits set by the \gls{LHC}. These calculations are consistent with ongoing dilepton 
		searches at the \gls{LHC} \citep{2013JHEP...08..091E}. To represent all the relevant interactions, we included the processes 
		$f\bar{f} \rightarrow \sum_n (\gamma^*/Z^*)_n \rightarrow F\bar{F}$ 
		where $f(F)$ can be any initial (final) state fermion.  
		
		The \gls{KK} bosons mediate processes between fermions \citep{2007AIPC..903..261D}. The \gls{KK} excitation masses are dependent on the extra 
		dimension size $R$ through the relation $m^* \equiv R^{-1}$, where $m^*$ is the mass of the \gls{KK} excitation. Thus, for the \gls{LKP} $n^{th}$ \gls{KK} 
		excitation masses $m_{Z^*}^{(n)}$ and $m_{\gamma^*}^{(n)}$ are given by,
		\begin{eqnarray}
			m_{Z^*}^{(n)} &=& \sqrt{m_{Z*}^{2}+(n\cdot m^*)^2}\\
			m_{\gamma^*}^{(n)} &=& n\cdot m^*
		\end{eqnarray} 
		
		We performed our calculations at logarithmically spaced energies starting from the \gls{LHC} energy, 14$\,$ TeV, to the maximum astrophysical energies, 
		$\sim 10^8\,$ TeV. We report the computed cross sections $\sigma(E)$ in figure \ref{fig:xsection} where $E$ is the beam energy. A larger cross section 
		indicates a higher probability that fermionic processes will be mediated by a \gls{KK} boson. All simulated values were for a 95\% confidence interval.
		
	\subsection{GRB Dynamics}
	\label{sec:dynamics}
		The excitation rate per unit volume is a Lorentz-invariant quantity and is dependent on the cross section for \gls{KK} excitation and the number and energy of 
		the protons in the outflow. In \gls{GRB}s, for these \gls{KK} excitations to be appreciable, they must proceed quickly with respect to the time-scales of the bursts. 
		The gamma-ray flux from \gls{GRB}s may last anywhere from two to several hundreds of seconds, with a few notable exceptions \citep{1993ApJ...413L.101K}. 
		In the previous section, we calculated the probability that a collision will give a \gls{KK} excitation. In this section, we describe the energy distribution of protons in 
		the outflow. 
		
		In our model, the central engine is spherical and its mass is concentrated in baryons. Since \gls{GRB}s preferentially occur in areas with low metallicity 
		\citep{2006ApJ...637..914W}, we assume a flow composed of free neutrons and protons. For long \gls{GRB}s, the progenitor object is believed to be a massive 
		star collapsing into a blackhole, so that the initial size is on the order of the Schwarschild radius $R_0 \gtrsim 2GM_0/c^2 \sim 10^3-10^5\;\rm m$. The energy 
		of the burst is related to its mass by $M_0 = E_0/\Gamma c^2$, where isotropic energy output of a \gls{GRB} can reach up to $10^{54}\;\rm ergs$ 
		\citep{2006RPPh...69.2259M}. The energy imparted to this mass $M_0 \ll E_0/c^2$, within the initial radius $R_0$, forces a relativistic expansion of the material. 
		The relativistic outflows of \gls{GRB}s have a coasting Lorentz factor on the order of $\Gamma_0 \gtrsim 100 - 1000$. An average coasting Lorentz factor of $100$ 
		and an isotropic energy output of $10^{54} \;\rm ergs$ would suggest an outflow mass with $M_0 \sim 10^{-3} \sm$. The coasting Lorentz factor is reached at a 
		saturation radius $R_s$. The Lorentz factor as a function of distance from the central engine $R$ is therefore
		\begin{equation} 
			\Gamma(r) = \left\{ 	\begin{array}{l l}
	    							R/R_0 & \quad \textrm{if $R \leq R_{sat}$}\\
    								\Gamma_0 & \quad \textrm{if $R \geq R_{sat}$}
  							\end{array} \right. ,
		\end{equation}
		where saturation radius is defined by $R_s \simeq r_0\Gamma_0$.
		
		As the fireball expands into a relativistic outflow, the initial thermal energy of the fireball is progressively converted to kinetic energy. From the observers frame, 
		the outflow organizes into a radial shell of width $\Delta R \sim R_0$ with the mass of the outflow given by $M = Nm_b$, where $N$ is the number of baryons 
		and $m_b$ is the baryon mass. We assume that protons and neutron have approximately the same mass, hence $m_b \approx m_p$. The comoving proton 
		density is given by,
		\begin{eqnarray}
			n'_p &\approx& \frac{E_0x}{4/3\pi R_0^3c^2m_b \Gamma}
		\end{eqnarray}
		where $x$ is the proton fraction, i.e. $x = n_p/n_b$, which we assume to be $0.5$.
		
		As the relativistic outflow expands into the \gls{ISM}, it produces shocks which inject baryons from the \gls{ISM} into the shell. \gls{GRB}s generally occur in 
		dense star forming regions, therefore we assume that the \gls{ISM} has a number density of $n_{ISM}$ = 1 cm$^{-1}$.  The shell will continue to propagate 
		with Lorentz factor $\Gamma_0$ in the coasting phase until it reaches the spreading phase at $r_S$.  In the spreading phase, the shell expands with its local 
		sound speed due to the velocity difference within the shell. The shell continues to propagate and expand, reaching the deceleration radius $r_D$, it begins to 
		lose kinetic energy to the cold \gls{ISM}. At which point, 
		\B
			r_D = \frac{3E_0}{4\pi n_{ISM}m_bc^2\Gamma_0^2}.
		\E
		
		Internal collisions occur when a shell catches up to another shell. Consider a system where each of the shells ejected from the central object have random 
		mass-energy injections and dimensionless velocity $\beta$, where $\Gamma = 1/ \sqrt{1-\beta^2}$. This phenomenon produces a number of internal collisions 
		between these shells. 
		
		Consider a faster shell ejected sometime after a slower shell, $\Delta T$. The faster shell, henceforth denoted by the subscript $f$, will sweep into the slower shell, 
		henceforth denoted by the subscript $s$. The faster shell injects energy into the slower shell, thus it will be called the injective shell. The slower shell receives the 
		injection of energy in the form of baryons. The collision creates a shock between the two shells, similar to the shock created between the shell and the \gls{ISM}. 
		The radius at which these two shells collide is 
		\B
			r_{coll} \simeq 2\Gamma_s^2c(\Delta T)/(1-\Gamma_s/\Gamma_f),
		\E
		where $\Gamma_f$ and $\Gamma_s$ are the Lorentz factors of the fast and slow shell respectively. Notice that in the limiting case where 
		$\Gamma_f\gg\Gamma_s$ that this reduces to $r_{coll}\simeq 2\Gamma_s^2(c\Delta T)$. For a shells with $\Gamma_f = 300$ and $\Gamma_s = 1$, 
		$r_{coll} \sim 10^6 \;\rm m$ if the fast shell is ejected $10 \;\rm ms$ after the slow shell.
		
	\subsection{Excitation Rate and Optical Depth}
	\label{sec:collisions}
	
		\gls{KK} excitation may occur in shocked regions as the outflow collides with either the \gls{ISM} or a slower shell. The shocks facilitate diffusive and stochastic 
		acceleration of the baryons, which results in a power-law energy distribution $n_p(E)\propto (E/E_0)^{-\alpha}$, where the spectral index is $\alpha\approx 2$. 
		We take the width of the shocked region to be $\Delta R \sim R_0$.
		
\begin{figure}[h!]
		\label{fig:kkoptdep}
		\center{
			\includegraphics[scale=0.3]{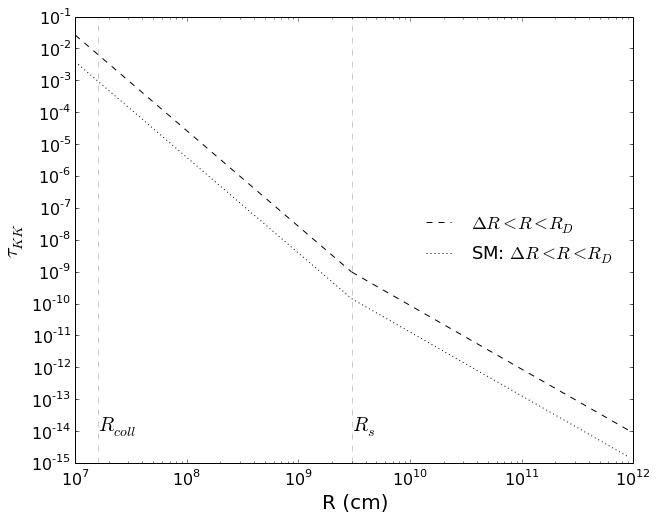}
			\includegraphics[scale=0.3]{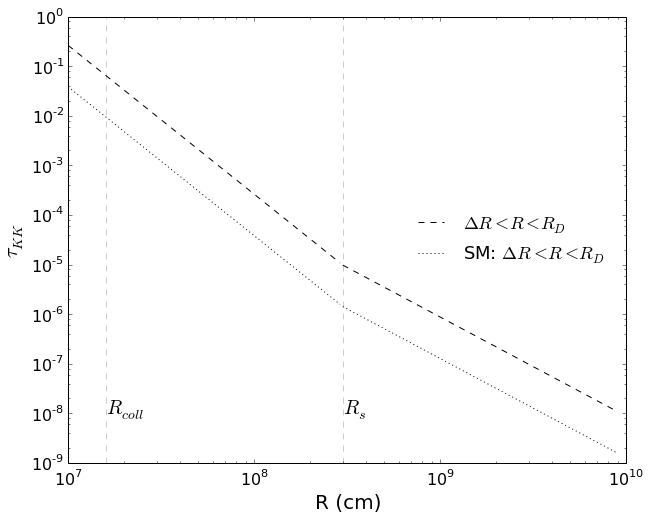}
			\includegraphics[scale=0.3]{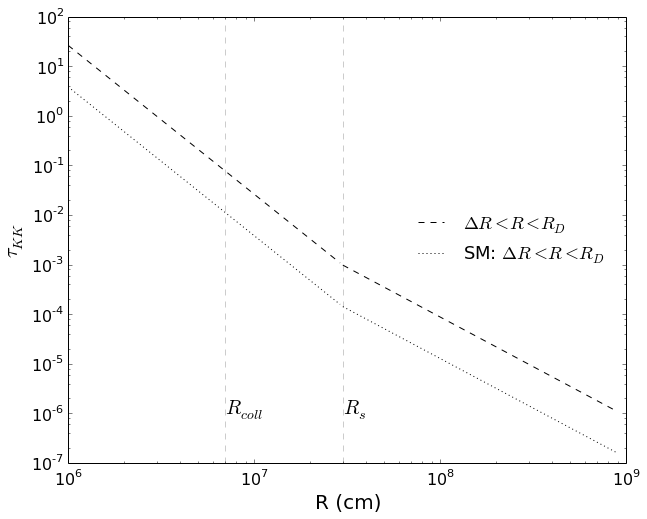}
			\caption{Energy integrated production optical depth for \gls{KK} excitations for various parameters. The optical depth indicates the likelihood that a 
				particle will undergo collisions resulting \gls{KK} excitation while traveling across the width of a shell. The plot is a function of the distance between 
				the shell and the central engine. In the case of a purely \gls{SM} regime, this is equivalent to the production optical depth \gls{SM} bosons $\gamma$ 
				or $Z^0$. The shell density is $10^{33}$ cm$^{-3}$ with a energy distribution that follows a power law with index $\alpha=-2$. The width of the shell is 
				$10^7$ cm. The Lorentz factor is 300. In the upper left panel, we assume multiple shells expanding with particle density $n_0$ on the order of 
				$10^{33}$ cm$^{-3}$. The upper right panel assume that the shell has slowed and the Lorentz factor is 30. Finally, the lower panel assume that the 
				burst outputs a single shell with $10^{35}$ cm$^{-3}$ and Lorentz factor is $30$.}
		}
	\end{figure}

		Once the protons have been accelerated above the resonant energy, collisions may produce \gls{KK} excitations. To quantify the importance of 
		\gls{KK}-mediated processes, we compute an excitation optical depth 
		\B
			\tau_{KK}\approx\frac{n_b(r)\langle\sigma_{KK}(r)\rangle\Delta R}{\Gamma}
		\E
		as a function the distance $r$ where the shell smashes into either a slower shell or the \gls{ISM}. Here we take the region where significant \gls{KK} mediated 
		collisions occur to be about the shell width $\Delta R$, and $n_b$ is the baryon number density in this region that depends on both $r$ and the nature of the 
		medium that the shell runs into. To compute $\tau_{KK}$, we use a energy-averaged cross section
		\B
			\langle\sigma_{KK}(r)\rangle = \frac{\int{\sigma(E,r)}\phi(E,r)dE}{\int{\phi(E,r)dE}},
		\E
		where $\phi(E,r) = n_b (E,r) v\approx n_b (E,r) c$ is the energy-dependent proton flux in the highly relativistic flow. Here we assumed that $n_b(E,r)$ is 
		approximately spatially constant across $\Delta R$. We report $\tau_{KK}$ in figure \ref{fig:kkoptdep} for different scenarios.
		
		Finally, we calculate the production rate. The nuclear collision rate is given by, $\dot{n}_{coll}(E)\- =\-n_{p,1}(E)\sigma(E)n_{p,2}c/\Gamma_{rel}$. The protons 
		in $n_{p,2}$ are assumed to be stationary with respect to $n_{p,1}(E)$. This assumption is intuitive for shocks with the \gls{ISM}. While injections of protons 
		by the \gls{ISM} may be accelerated in these shocks, it is much more reasonable that they collide with the already accelerated protons. Thus, the production 
		rate is $\dot{n}_{KK}(E)\-= n_{p,s}(E)\sigma(E)n_{p,ISM}c/\Gamma_{rel}$, where $n_{p,s}$ and $n_{p,ISM}$ are the particle densities for the shell and \gls{ISM}, 
		respectively. In internal collisions, the assumption is that there is a fast shell and a slow shell. In the strongly shocked internal collision region, there may be 
		head-on and tail collisions. In the dynamics of Fermi acceleration, head-on collisions are much more likely \cite{1949PhRv...75.1169F}. If $\Gamma_{rel}\gg 1$, 
		then this simplifies to collisions with a stationary target. Thus, $\dot{n}_{KK}(E) = n_{p,f}(E)\sigma(E)n_{p,s}c/\Gamma_{rel}$, where $n_{p,f}$ and $n_{p,s}$ are 
		the particle densities for the fast and slow shell, respectively.
		
\section{Results}
\label{sec:results}
	
\begin{figure}[h!]
		\center{
			\includegraphics[scale=0.5]{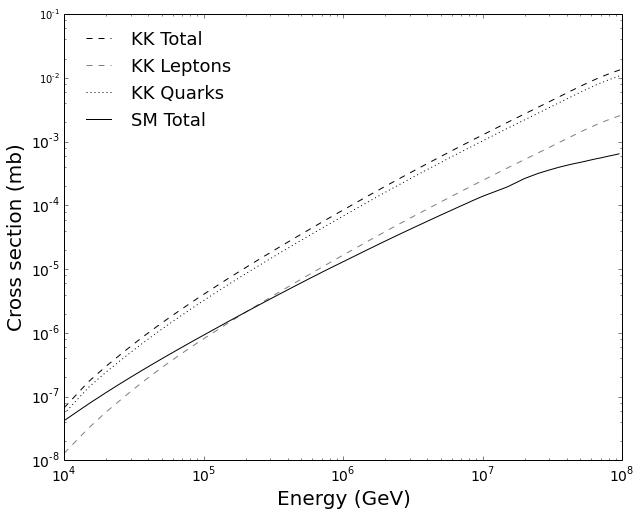}
			\caption{Pythia8 generated cross sections for KK excitations. All values are within a 95\% confidence interval.}
		}
		\label{fig:xsection}
	\end{figure}

	Our calculation gives the number of events as a function of energy, and a total cross section. As the collision energy rises, the \gls{KK} excitation cross section 
	increases and diverges further from its \gls{SM} counterpart due to the larger influence from higher order processes \cite{2008NuPhB.797....1D}. The cross section 
	for producing quark antiquark pairs is approximately an order of magnitude greater than for producing leptons. All cross sections are within a 95\% confidence 
	interval. These Monte Carlo generated cross sections are shown in figure \ref{fig:xsection}. 
	
\begin{figure}[h!]
		\center{
			\includegraphics[scale=0.5]{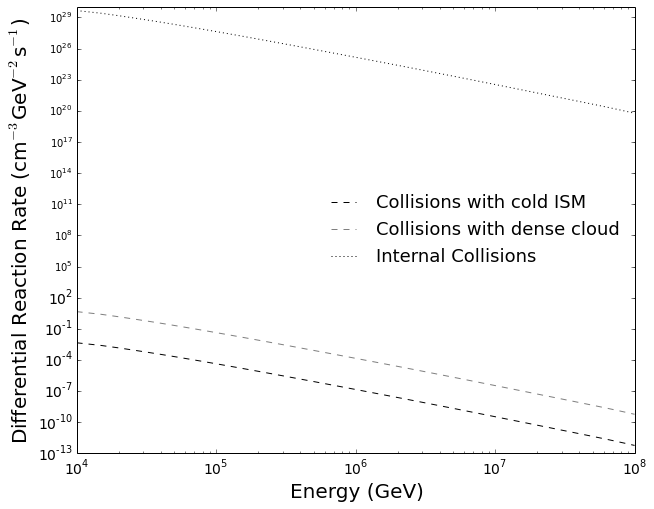}
			\caption{Differential reaction rate for \gls{KK} excitations. The densities are 1 cm$^{-3}$ and 1000 cm$^{-3}$ for the \gls{ISM} and dense molecular cloud, 
			respectively. The isotropic energy output for the baryons associated with each shell is $10^{52}$ ergs and the initial radius is $10^7$ cm. The incident 
			particle density $n_0$ was on the order of $10^{33}$ cm$^{-3}$. For internal collisions, we assume that the slow shell has slowed such that 
			$\Gamma_{rel} \sim \Gamma_f$.}
		}
		\label{fig:rate}
	\end{figure}

	We calculated production rates when a shell smashes into the \gls{ISM}, a dense molecular cloud, and a slower shell. For the shells, we used a mass of $10^{-3}\;\sm$, 
	which is consistent with the mass estimates in section \ref{sec:dynamics}. We assumed that Fermi-type acceleration generated a power-law spectrum for the 
	differential particle densities in the shocks with energies from 100 to $10^8 \;\rm GeV$ and a Lorentz factor of 300. For the internal collisions, we report the reaction 
	rates between two shells with the same mass, a fast shell with a Lorentz factor of 300, and a slower target shell with a Lorentz factor of 1. The \gls{KK} excitation rates 
	are reported in figure \ref{fig:rate}. The total reaction rates in cm$^{-3}$s$^{-1}$ were 53, 5300, and 5.3$\times10^{33}$ for \gls{ISM} case, dense cloud case, and 
	internal collisions.
	
\section{Conclusions and Observational Implications}
\label{sec:conclusion}
	
	Our results indicate that \gls{KK} excitations could be significant as a \gls{GRB} fireball shell collides with another medium, especially since \gls{KK} mediated 
	processes enjoy vastly boosted cross-sections over their \gls{SM} equivalents over the relevant energy range. In particular, we found that collisions with the 
	\gls{ISM} and a dense cloud would produce \gls{KK} excitations, but at a much slower rate than internal collisions. Despite the increase in the cross-section with 
	particle energy, the excitation rate decreases as there are fewer protons at higher energies. 
	
	Reference \cite{2010JHEP...09..025B} found a characteristic peak and valley in the \gls{KK} excitation rate as a function of beam energy. The suppression and boosting 
	of the cross section that leads to these features is consistent with our simulations. The \gls{LKP} should be identifiable by the characteristic suppression of the number 
	of events near its resonant rest mass energy ($1.59 \rm \, TeV$ in our work). This would in turn lead to a decrease in gamma ray output from a \gls{GRB} near and 
	above this energy scale. Such a dark matter production signal does not conflict with any observational constraints, and would account for the signals gamma-ray 
	excess from the Fermi and Pamela experiments. In the near future, improvements to the current generation of TeV instruments, the newly operational \gls{HAWC} 
	experiment, and the planned \gls{CTA} experiment aim to observe \gls{GRB}s in the TeV regime. In particular, \gls{CTA} is expected to revolutionize the study of 
	cosmic particle accelerators, and may provide further evidence of \gls{DM} production.
	
	We assumed that \gls{GRB} fireballs are baryon dominated, however other theories assume that the outflow is composed of leptons, primarily electrons. The resulting 
	spectrum is then from one of or a combination of synchrotron, inverse Compton, and synchrotron self-Compton. The benefits of these leptonic models include their 
	relative simplicity, and the numerous adjustable parameters for fitting purposes. On the other hand, the complexity of hadronic models has limited their use in 
	simulations as the codes must keep track of all the particles from particle showers and determine their contribution to the spectrum. Despite these difficulties, there 
	has been some recent work attempting to explain observations with hadronic flows \citep{agm09, bel10}. Using similar methods, we hope to compute the predicted 
	\gls{GRB} spectrum taking \gls{KK} excitations into account. 
	
	In the future, we also intend to explore the possibility that \gls{SNR} may be significant \gls{KK} production environments. As a source of galactic cosmic 
	rays,  \gls{SNR}s also create shocks and accelerate protons. \gls{SNR}s live longer than \gls{GRB}s and have been observed in the TeV range 
	\citep{2008A&A...481..401A}. Furthermore, hadronic, photopion models have already been shown to fit the gamma-ray spectrum from \gls{SNR}s 
	\citep{2013Sci...339..807A}.

\section{Acknowledgements}
	The authors acknowledge the support from St.~Mary's College of Maryland faculty development grant. We also appreciate insightful discussions with P. Meszaros and 
	O. Blaes.

\section*{References}
	\bibliographystyle{elsarticle-num} 
	\bibliography{kk_ref.bib}

\end{document}